\documentclass[11pt]{article}

\usepackage{amsmath,amssymb,url}
\usepackage[english]{babel}

\begin{document}
\title{Jetted Narrow-Line Seyfert 1 Galaxies \& Co.: where do we stand?}

\author{Luigi Foschini\footnote{INAF Brera Astronomical Observatory, 23807 Merate (Italy); email: {\tt luigi.foschini@inaf.it}}}
\date{26 August 2020}
\maketitle

\begin{abstract}
The discovery in 2008 of high-energy gamma-rays from Narrow-Line Seyfert 1 Galaxies (NLS1s) made it clear that there were active galactic nuclei (AGN) other than blazars and radio galaxies that can eject powerful relativistic jets. In addition to NLS1s, the great performance of the {\it Fermi} Large Area Telescope made it possible to discover MeV-GeV photons emitted from more classes of AGN, like Seyferts, Compact Steep Spectrum Gigahertz Peaked Sources (CSS/GPS), and disk-hosted radio galaxies. Although observations indicate a variety of objects, their physical characteristics point to a central engine powered by a relatively small-mass black hole (but, obviously, there are interpretations against this view). This essay critically reviews the literature published on these topics during the last eight years and analyses the perspectives for the forthcoming years. -- {\it Keywords:} Relativistic Jets; Active Galactic Nuclei; Seyfert; Blazar; High-Energy Gamma Rays
\end{abstract}


\section{Introduction: the status before 2012}
High-energy gamma rays from jetted narrow-line Seyfert 1 galaxies (NLS1) were detected for the first time in $2008$ by the {\it Fermi} Large Area Telescope (LAT), after the early months of operations \cite{ABDO1,ABDO2,ABDO3,FOSCHINI1}. In 2012, I wrote an extensive review on this topic, dealing with the history before and after the discovery of the gamma-ray emission, and I refer the reader to that article \cite{FOSCHINI2}. This review begins where I ended the previous essay. The main open questions in 2012 were:

\begin{enumerate}
\item The nature of jetted NLS1s: the place of these objects in the family of jetted active galactic nuclei (AGN) was still to be understood. A certain degree of similarity with flat-spectrum radio quasars (FSRQs) was apparent since the early studies \cite{ABDO1,ABDO2,ABDO3}, but other observational differences (relatively low observed luminosities, compact radio morphology, narrowness of permitted emission lines, different host galaxy...) did not permit to establish if NLS1s and FSRQs are drawn from the same population or if there are intrinsic differences. One critical issue was the estimate of the mass of the central black hole.    
\item The parent population: five NLS1s with the jet viewed at small angles (bulk Lorentz factor $\Gamma \sim 10$), require at least $5\times 2\Gamma^2 \sim 1000$ objects with the jet viewed at large angles. However, in 2012, only four cases of jetted NLS1s with large viewing angles were known. What and where are the missing objects?
\end{enumerate}

Although, NLS1s were the first non-blazar AGN to be detected at high-energy gamma rays, other types of AGN were later discovered to be gamma-ray emitters (Seyferts, Compact Steep Spectrum Gigahertz Peaked Sources CSS/GPS, and disk-hosted radio galaxies). Therefore, this review mostly deals with NLS1s, but also includes other AGN displaying relativistic jets as they are all member of the set of jetted AGN powered by small-mass black holes.

\section{The most important discovery}
Looking back to these eight years, I think that the most important discovery was made in 2018 by L\"ahteenm\"aki et al. \cite{LAHTEENMAKI2018} with the 14-m single-dish radiotelescope at Mets\"ahovi (Finland) operating at 37~GHz. In the framework of a large program to monitor different samples of jetted AGN, L\"ahteenm\"aki  et al. discovered strong outbursts from NLS1 formerly classified radio quiet or even silent! Such strong outbursts (flux densities at Jy level) at high radio frequency (37~GHz) can be due only to a relativistic jet, which is at odds with the classification radio quiet or silent. I would like to stress that the point worth noting in the work by L\"ahteenm\"aki et al. is the change radio quiet/loud many times per year. There are known examples of radio quiet AGN with relics of a past jet activity (e.g. \cite{MARECKI,CONGIU2017,CHEN2020}), but, in these cases, their jet was switched off as a consequence of the cosmological evolution. Today, they can safely be labelled as radio quiet. Quite the opposite, the NLS1s discovered by L\"ahteenm\"aki et al. change their state many times per year, making the radio loud/quiet label useless. As already pointed out by Padovani \cite{PADOVANI2017}, it is time to move toward a more physical classification, as jetted or non-jetted AGN.

Follow-up programs of these NLS1s have been activated, so that it will be interesting to read forthcoming works to better understand what can trigger such violent and erratic activity. 

\section{The mass of the central black hole}
\label{mass}
Non-jetted NLS1s are known to be AGN with small black hole masses and high accretion rates, which in turn is confirmed by the morphology of their host galaxy, generally spirals with pseudobulges or bars \cite{PETERSON1,PETERSON2}. Blazars (FSRQs and BL Lac Objects, plus radiogalaxies, their parent population) are massive AGN hosted by giant ellipticals \cite{UP}. These two paradigms clashed when the detection of high-energy gamma rays from NLS1s confirmed the presence of powerful relativistic jets in this type of AGN. Researchers divided into two communities centered on the estimate of the black hole mass: small ($10^{6-8}M_{\odot}$) vs high ($10^{8-9}M_{\odot}$) masses. This is a key question in these studies, with multiple implications that I will address later in this review. In this section, I focus only on the mass estimate. 

As non-jetted NLS1s are known to have black hole masses in the $10^{6-8}M_{\odot}$ range, the small-mass paradigm was favored already in the early papers on the discovery of high-energy gamma ray emission, suggesting that jetted NLS1s could be the low-mass tail of the distribution of FSRQs \cite{ABDO1,ABDO2,ABDO3}. These essays also studied the possible effects of the radiation pressure (according to Marconi et al. theory \cite{MARCONI}), and the geometry of the broad-line region (BLR, according to the Decarli et al. model \cite{DECARLI}), but just for the sake of completeness, without undermining the small-mass paradigm.  

The rift in the researcher community was set in 2013 by Calderone et al. \cite{CALDERONE}. They proposed a detailed analysis of the infrared-to-ultraviolet spectra of a sample of jetted NLS1s, concluding that the mass estimates should be a factor six greater than the virial ones ($0.8$ dex). Their method begins with measuring the disk luminosity via the main emission lines (e.g. H$\beta$, MgII). As the black body is a self-similar function, setting the disk luminosity $L_{\rm disk}$ results also in fixing its value at the peak $\nu_{\rm peak}L_{\nu,\rm peak}=0.5L_{\rm disk}$~erg~s$^{-1}$ (Eq. A10 in \cite{CALDERONE}). Then, by using spectra from the \emph{Sloan Digital Sky Survey} (SDSS) and the standard accretion disk model by Shakura \& Sunyaev, it is possible to constrain the peak frequency and thus the black hole mass. Calderone et al. also performed an estimate by adding further constraints from ultraviolet data of the \emph{Galaxy Evolution Explorer} (\emph{GALEX}) and by subtracting the jet contribution by using infrared data from \emph{Wide-Field Infrared Survey Explorer} (\emph{WISE}). The host galaxy contribution was also subtracted by adopting templates from \cite{MANNUCCI}. 

The reasons of this significant discrepancy with the virial estimates could be due to many assumptions not suitable for the specific context. Just to cite a few: the use of Shakura-Sunyaev disk, without taking into account possible and more likely alternatives; the viewing angle is set to $\theta \sim 30^{\circ}$ (Sect. 4.1, item number IV in \cite{CALDERONE}), which is not consistent with what expected for relativistically beamed sources ($\theta \lesssim 10^{\circ}$); the jet emission is evaluated from W3 and W4 \emph{WISE} magnitudes, but infrared emission from NLS1s is significantly affected by star formation \cite{CACCIANIGA}; there was no control sample, made with objects with well-known and reliable masses measured by means of reverberation mapping; the cherry picking of the objects (the original sample was made of 23 NLS1s, but 6 were excluded because data did not agree with the model, Sect. 5 in \cite{CALDERONE}). However, in my opinion, the critical error was to set the radius of the innermost stable orbit equal to six times the gravitational radius ($R_{\rm in}=6R_{\rm g}$, Sect. 4.1, item number III in \cite{CALDERONE}). As the high-frequency emission comes from the innermost regions of the accretion disk, by setting such large value for the inner radius is equivalent to cutting off contributions from higher frequencies, where the accretion disk around small mass black holes has its peak emission. This is also confirmed by Castell\'o-Mor et al. \cite{NCM}, who performed a similar study on a sample of well-known objects with masses calculated via reverberation mapping, but they left the inner radius of the accretion disk free to change depending on the spin of the black hole. In this case, the results from the accretion disk fitting were well in agreement with the virial estimates. 

A rule-of-thumb estimate of how fixing the innermost stable orbit to $6r_{\rm g}$ affects the peak temperature of the accretion disk can be calculated -- for example -- by using Eq.~(7.20) of \cite{FKR}:

\begin{equation}
T(r) \sim 2.3\times 10^{4} \sqrt{\frac{600 r_{\rm g}}{r}} \left(\frac{L}{L_{\rm Edd}}\right)^{1/4} \left(\frac{M}{10^{8}M_{\odot}}\right)^{-1/4} \, {\rm [K]}
\end{equation}

It is possible to note how the product by the constant and the radius-dependent term ($2.3\times 10^{4}\sqrt{600 r_{\rm g}/r}$~K) change by setting $r=6r_{\rm g}$ ($\rightarrow 2.3\times 10^{5}\, {\rm K} \sim 4.8\times 10^{15}$~Hz) or $r\sim r_{\rm g}$ ($\rightarrow 5.6\times 10^{5}\, {\rm K} \sim 1.2\times 10^{16}$~Hz), the latter being the case of a prograde Kerr black hole maximally rotating. A difference by factor $2.5$ implies a difference by a factor $2.5^{4}\sim 39$ in mass, when keeping the accretion luminosity constant. Therefore, it follows that truncating the innermost part of the accretion disk will result in greater masses. 

The fit of the standard accretion disk model was adopted also by Ghisellini et al. in their one-zone radiative leptonic modeling of the spectral energy distribution (SED) of blazars \cite{GHISELLINI2010,GHISELLINI2015}. In this case, the SED modeling makes it possible to perform a better subtraction of the jet emission, but the mass overestimate still holds, although the discrepancy is smaller than Calderone et al.'s work: about $0.2$~dex vs $0.8$~dex, with larger deviations toward small masses (see Fig. 6 in \cite{GHISELLINI2015}). 

\subsection{Case study: PKS~$2004-447$}
A different approach to solve the mass conundrum was to use optical spectropolarimetry \cite{BALDI}. The case for polarization in NLS1s is weaker than the well-known case of Type 2 AGN, where the scattering is due to the obscuring torus. In NLS1s, the disk and BLR are directly observed, as proved by the FeII bumps in the optical spectra. Furthermore, in jetted NLS1s viewed with small angles $\lesssim 10^{\circ}$, the polarized contributions from opposite sides of the BLR cancel each other: a viewing angle of $0^{\circ}$ and a symmetrical BLR would result in $0$\% net polarization. 
Baldi et al. \cite{BALDI} observed PKS~$2004-447$ ($z=0.24$) at European Southern Observatory (ESO) Very Large Telescope (VLT) and found almost nothing ($P=0.03\pm 0.02$\% across the whole spectrum, $5404-7254$\,\AA), but with some hint ($P=0.0076\pm0.0024$\%) around the H$\alpha$ emission line ($6534-6594$\,\AA). The statistics was very poor, and the authors themselves admitted that they cannot exclude it was just the residual instrumental polarization. Moreover, the observation was not flux calibrated, so it was not possible to evaluate the contribution from the jet: according to the SED, the optical emission should be dominated by the synchrotron emission \cite{ABDO3,ORIENTI2015}, which implies a significant degree of polarization. Focusing on the sides of the H$\alpha$ line, the polarized signal improves a little bit ($P=0.066\pm0.023$\% in the range $6424-6504$\,\AA; $P=0.070\pm0.023$\% in the range $6624-6704$\,\AA). Baldi et al. inferred a FWHM(H$\alpha$)$=9000\pm 2300$~km/s, from which they calculated a mass of the central black hole of $\sim 6\times 10^{8}M_{\odot}$ \cite{BALDI}. In 2001, Oshlack et al. estimated the virial mass to be $\sim 5.4\times 10^{6}M_{\odot}$ \cite{OSHLACK}, but after a $\gamma$-ray ourburst in 2019, Berton et al. \cite{BERTON2019} found an error in the flux value of Oshlack et al.'s spectrum, which was lower by two orders of magnitude than the original observation by Drinkwater et al. \cite{DRINKWATER}. Therefore, the revised virial mass was $\sim 7.0\times 10^{7}M_{\odot}$ \cite{FOSCHINI2015}, still one order of magnitude smaller than Baldi et al.'s value, based on poor statistics.

A soft X-ray excess was noted already by Gallo et al. \cite{GALLO2006} and confirmed by Foschini et al. \cite{FOSCHINI2009}. The 2004 {\it XMM-Newton} observation was fit by a thermal Comptonization model with a temperature of the seed photons equal to $kT=66\pm 17$~eV \cite{GALLO2006}: if due to the Wien tail of the multicolor black body of the accretion disk, it would be consistent with a small black hole mass. However, this soft excess was no longer present in more recent observations performed in 2012 when the source was in a lower flux state \cite{ORIENTI2015,KREIKENBOHM2016}. If the soft excess is due to the accretion disk, one expects that it emerges as the jet continuum decreases, while the opposite was observed. This, together with some significant variability ($RMS=16\pm4$\% in the $0.2-1$~keV band \cite{FOSCHINI2009}), points to a different origin of this feature for PKS~$2004-447$, perhaps a tail of the synchrotron emission. 

To summarize, the search for polarization resulted in a hint of some signal, but the statistics is not enough to draw scientifically sound conclusions. More observations are needed to understand if it is a signal from the source or an artifact due to the residual instrument polarization. In this case, the soft excess is not useful to estimate the black hole mass, because it is likely not due to the accretion disk. The virial estimate remains the only reliable value found to date.

\subsection{Case study: IC~310}
This object is a Seyfert~2 galaxy at $z=0.019$, with prominent H$\alpha$+[NII] lines, hosted by a SA galaxy. It was detected at high-energy gamma rays by {\it Fermi}/LAT, with a rather hard spectrum and significant emission above 100~GeV \cite{NERONOV2010}, later confirmed by the {\it MAGIC} Cherenkov telescope \cite{ALEKSIC2010}. Its radio structure displays a one-side jet, indicating significant beaming, although it was not possible to set tight constraints to the viewing angle ($\theta \lesssim 38^{\circ}$, \cite{KADLER2012}). In 2014, the {\it MAGIC} Collaboration reported the detection of unusual variability at very high energies, with flux doubling time scales of $4.8$~minutes \cite{ALEKSIC2014}. The mass of the central black hole was estimated to be $3_{-2}^{+3}\times 10^{8}M_{\odot}$ by using the bulge velocity dispersion and the fundamental plane, which implies a light crossing time (that is also the minimum scale for variability) of $\Delta t = 23_{-15}^{+34}$~min \cite{ALEKSIC2014}. To make the measured value of $4.8$~min consistent with the light crossing time of $23$~min, one can invoke a proper Doppler factor, but the authors excluded this option on the basis of the jet viewing angle and the need to avoid pair creation to detect GeV photons. They concluded that the emission region should be sub-horizon \cite{ALEKSIC2014}. 

However, one year later, a new estimate of the mass of the central black hole ($3\times 10^{7}M_{\odot}$, \cite{BERTON2015}) solved the conundrum. A smaller mass implies a smaller light crossing time, which is now one order of magnitude smaller and consistent with the observed $\sim 5$~min variability. The {\it MAGIC} Collaboration conceded that the smaller mass weakens the sub-horizon argument, but insisted that the opacity still points to the initial hypothesis \cite{AHNEN2017}. However, other researchers have already pointed out how unlikely is a sub-horizon emission from IC~310 on the basis of the observed gamma-ray luminosity, which was one order of magnitude greater than the maximum allowed for a magnetospheric origin \cite{AHARONIAN2017,KATSOULAKOS2018}. They recognized that a smaller mass of the central black hole would solve many problems, except that of the accretion disk. They all assumed an advection-dominated accretion flow (ADAF), although it is not easy to understand on what basis: perhaps, as IC~310 was indicated to be similar to a Fanaroff-Riley type I/BL Lac Object jetted AGN, it was taken for granted that the accretion power was weak and unable to support other types of disk. However, the optical spectrum from SDSS displays a prominent H$\alpha$+[NII] line complex, plus more weak lines: these evident lines indicated the presence of a radiatively efficient disk (the ADAF spectrum displays a featureless continuum with a peak at infrared frequencies \cite{BECKERT2002}). From the SDSS spectrum it is possible to roughly estimate the luminosity of the H$\alpha$ line to be $L_{{\rm H}\alpha}\sim 1.7\times 10^{43}$~erg~s$^{-1}$. By using the relationships linking the line luminosity to the BLR one, and, in turn, to the accretion disk (e.g. \cite{GHISELLINI2015}), it results $L_{\rm disk}\sim 10L_{\rm BLR}\sim 10\cdot (556/77)\cdot L_{{\rm H}\alpha}\sim 10^{45}$~erg~s$^{-1}$. For a $3\times 10^{7}M_{\odot}$ black hole, this means an Eddington ratio $\sim 0.31$, clearly not the one of an ADAF. Therefore, the most likely explanation of the short variability observed by {\it MAGIC} is a small black hole with high accretion, as indeed is expected for a Seyfert galaxy.

\subsection{Case study: 1H~$0323+342$}
The nearby gamma-ray NLS1 1H~$0323+342$ ($z=0.063$) is the only relativistically beamed object with an estimate of the mass of the central black hole via reverberation mapping: Wang et al. performed a campaign in 2012 with a $2.4$~m telescope located in Lijiang (China) \cite{WANG2016}. They found a reverberation delay of about two weeks in the H$\beta$ and FeII emission lines: $14.8_{-2.7}^{+3.9}$ days for H$\beta$, and $15.2_{-4.1}^{+7.4}$ days for FeII. From these values, they calculated a black hole mass of $3.4_{-0.6}^{+0.9}\times 10^{7}M_{\odot}$ \cite{WANG2016}.  This result was confirmed by Landt et al., who calculated the mass by using different methods and infrared, optical, and X-ray data: the resulting average value was $\sim 2\times 10^{7}M_{\odot}$ \cite{LANDT2017}. Also the virial method, with single-epoch spectrum, and the measurement of the dispersion of the emission line (which is less affected by the inclination and the accretion rate, \cite{COLLIN}), gives consistent result ($\sim 3.6\times 10^{7}M_{\odot}$, \cite{FOSCHINI2015}). The only discrepancies (about one order of magnitude) were found in two specific methods: black hole-bulge relationship ($1.6-4.0\times 10^{8}M_{\odot}$ \cite{LEONTAV2014}), excess variance and power spectrum density bend frequency at X-rays ($2.8-7.9\times 10^{6}M_{\odot}$, \cite{PAN2018}). In both cases, the explanation of the discrepancies could be a contamination of the jet emission: in the former case, it can decrease the photometric magnitude, thus implying a larger mass; in the latter, the Doppler reduced variability at X-rays can result in a smaller mass. It is also worth noting that the use of the black hole-bulge relationship in disk-hosted galaxies could be not appropriate at all \cite{BOHN2020}.

To conclude, this is the only jetted NLS1 with the central black hole mass measured by using the reverberation mapping.  Most of the methods adopted by many authors converged to that value. There are only two estimates in disagreement, because the jet contribution was not properly taken into account. This is a well-grounded result that can be adopted as a benchmark.

\subsection{The mass of the central black hole of a jetted AGN}
The best known method to estimate the mass of the central black hole of an AGN would be a reverberation mapping campaign and the calculation of the corresponding velocity-delay maps \cite{PETERSON2014}, but it is a complex and time consuming task. An acceptable compromise is the use of single-epoch spectra with proper calibration relationships. Among the different options, the use of the line dispersion $\sigma_{\rm line}$ yields better results than the FWHM one, because the $\sigma_{\rm line}$ is less affected by the inclination and the accretion rate \cite{PETERSON2004,COLLIN,DALLABONTA2020}. One would expect that for a given mathematical profile the two quantities are related, but observations indicate something different and I refer to the excellent works by \cite{PETERSON2004,COLLIN,DALLABONTA2020} for more detailed analyses. The key point is the measurement of $\sigma_{\rm line}$, which is less affected by the line wings.
In addition, for jetted AGN, the use of the line luminosity to estimate the radius of the BLR $R_{\rm BLR}$ is better than the continuum at $5100$\AA \, or any other continuum luminosity, because it is less affected by the jet boosted emission. That is, the virial mass is calculated according to:

\begin{equation}
M=f\frac{R_{\rm BLR}\sigma_{\rm line}^{2}}{G}
\end{equation}

where $G$ is the universal gravitational constant, $f$ is a dimensionless scale factor ($f=3.85-4.47$ according to \cite{COLLIN,WOO2015}), and the radius of the BLR is estimated according to \cite{GREENE}:

\begin{equation}
\log R_{\rm BLR} = 0.85 + 0.53 \log L(H\beta)
\end{equation}

where $R_{\rm BLR}$ is in units of 10 light days, and $L(H\beta)$ is in units of $10^{43}$~erg~s$^{-1}$. This is the most robust method to estimate the mass of the central black hole in a jetted AGN without making use of the reverberation mapping, because the effects of inclination, accretion rate, and jet continuum are minimized. By using this method, it was possible to estimate the mass of the central black hole of a sample of 42 jetted NLS1s with flat radio spectra: the resulting range was $\sim (5\times 10^{6}-3\times 10^{8})M_{\odot}$, with accretion luminosities in the range $\sim (0.01-0.49)L_{\rm Edd}$ \cite{FOSCHINI2015}. The same method applied to a sample of 18 steep-spectrum jetted NLS1s resulted in a similar mass and accretion luminosity ranges: $\sim (2\times 10^{6}-3\times 10^{8})M_{\odot}$, and $\sim (0.001-0.52)L_{\rm Edd}$ \cite{BERTON2015}.

\section{Host galaxy}
\label{hostg}
As the central supermassive black hole develop together with the host galaxy (see \cite{GRAHAM2016} for a review), the issue of the mass also led to the study of the host galaxies of jetted NLS1s. FSRQs, BL Lac Objects, and radiogalaxies are almost all hosted by giant dead elliptical galaxies \cite{UP}, while non-jetted NLS1s are generally hosted by spiral galaxies, often barred, and with pseudobulges \cite{MATHUR2012}. Then, the question was to understand if jetted NLS1s also have the same type of host galaxy. 

This research topic literally exploded during the latest years. Before 2012, there was only one case study, 1H~$0323+342$, and there was the doubt that an observed ring feature around the nucleus could be a spiral arm \cite{ZHOU2007} or the residual of a recent merger \cite{ANTON2008}, which in turn was later supported by another work \cite{LEONTAV2014}. In 2016, it was the turn of PKS~$2004-447$: VLT observations under excellent seeing ($0.40"-0.45"$) indicated a barred disk galaxy with a pseudobulge grown via secular processes \cite{KOTILAINEN2016}. In 2017, two independent works on the host galaxy of FBQS J$1644+2619$ were published: the first one reported a barred spiral SB0 \cite{OIGLE2017}, while the other claimed an elliptical host \cite{DAMMANDO2017}. The discrepancy seems to be due to the worse seeing during the observation of the second group ($0.9"$ vs $0.63"-0.75"$ of the first team), which could have blurred the disk features observed by the first team. 

Structures of interactions or recent mergers are often observed: PKS~$1502+036$ ($z=0.408$, elliptical bulge plus nearby ring, recent merger, \cite{DAMMANDO2018}), IRAS~$20181-2244$ ($z=0.185$, ongoing interaction between two galaxies, the NLS1 seems to be hosted by a disk galaxy, \cite{BERTON2019B}), TXS~$2116-077$ ($z=0.26$, ongoing merger, disk galaxy with pseudobulge, \cite{PALIYA2020,JARVELA2020}, reclassified as intermediate Seyfert-type active nucleus \cite{JARVELA2020}). In the case of SBS~$0846+513$ ($z=0.584$), observations with the Large Binocular Telescope were not sufficient to distinguish between the two main options \cite{HAMILTON2020}. 

In addition to these case studies, there are also the early surveys: J\"arvel\"a et al. \cite{JARVELA2018} studied in the $J$ filter a sample of 9 jetted NLS1s, being able to resolve 5 hosts, which in turn resulted to be disk galaxies with pseudobulges in all the five cases; four hosts also showed bars, and three cases have features of recent mergers. A larger sample, made of 29 jetted NLS1s (12 were also $\gamma$-ray emitters), was studied by Olgu\'in-Iglesias et al. \cite{OIGLE2020}, who found similar results. In addition, in a plot with nuclear vs bulge magnitude, jetted NLS1s are placed in the low-luminosity tail of FSRQs (Fig.~4 in \cite{OIGLE2020}), and also the Kormendy relationship shows clear deviations from the blazar region (Fig.~3 in \cite{OIGLE2020}). 

The main conclusion that can be drawn from these studies is that the host galaxy does not affect the jet formation: there is no preferred host type, thus implying that it does not matter. Before NLS1s, almost all the blazars were hosted by giant elliptical galaxies, so that one could have the reasonable doubt that there was some link with the jet formation. However, as jetted NLS1s display a variety of hosts (spirals, disk, ellipticals, interacting, with or without either pseudobulges, bars, or signs of recent mergers), it is now clear that there is no link between the jet formation and the host galaxy. This confirms what Blandford told in 1978 during the discussion of his seminal talk on blazars at the renowned Pittsburgh conference \cite{BLANDFORD1978}: as the AGN is confined to the central parsec, the host does not matter. This does not exclude that the engine could imprint some feedback on the galaxy (see Sect.~\ref{outflo}).

\section{Jetted NLS1s as the low-luminosity tail of the FSRQs distribution}
There are many observational features suggesting some similarity between NLS1s and quasars. Already at the end of nineties, it was suggested that NLS1s could be young AGN at the early stage of evolution \cite{GRUPE1999,GRUPE2000}, the low-$z$ analogues of high-$z$ quasars, with gas-rich hosts rejuvenated by recent mergers \cite{MATHUR2000,MATHUR2000B}. In the case of jetted NLS1s, already the early observations suggested the same. 1H~$0323+342$ showed a clear spectral variability at X-rays, which was interpreted in the framework of the jet-disk connection: when the jet is at low activity, the X-ray spectrum is dominated by the thermal Comptonization of the accretion disk corona ($\Gamma_{\rm 0.3-10\, keV}\sim 2$); when the jet emission overwhelms the disk one, a hard tail ($\Gamma_{\rm 3-10\, keV}\sim 1.4$, break energy $\sim 3$~keV) emerges \cite{FOSCHINI2009,FOSCHINI2} (particularly, see Fig.~1, left panel, in \cite{FOSCHINI2}; later confirmed by \cite{PALIYA2014}). A similar behavior was observed at hard X-rays: {\it INTEGRAL}/IBIS recorded a soft spectrum at low flux, while {\it Swift}/BAT reported a hard spectrum at high flux \cite{FOSCHINI2009}. 

This reminds the behavior of the archetypical FSRQ 3C~273: observations with {\it BeppoSAX} between $1996$ and $2001$ revealed the presence of a soft thermal component (Seyfert-like) and a hard continuum extending up to $\sim 200$~keV (jet) \cite{GRANDI2004}. When the jet dominates, the Seyfert-like component is overwhelmed, while the latter emerges as the jet activity decreases \cite{GRANDI2004}. This was also later confirmed by observations with {\it XMM-Newton} and {\it INTEGRAL} \cite{FOSCHINI2006,BIANCHIN2008}. 

With the detection of high-energy gamma rays, there was one more opportunity to test this similarity. Already since the early {\it Fermi}/LAT observations, it was suggested that jetted NLS1s could be the low-mass tail of the FSRQs distribution, although it was thought that the main driver to link the two populations was the Eddington ratio \cite{ABDO1,ABDO2,ABDO3}. The proof of the NLS1-FSRQ link was later established by following different approaches: 

\begin{itemize}
\item by comparing observational and physical properties \cite{FOSCHINI2015}; 
\item by calculating the luminosity function \cite{BERTON2016} (particularly, see Fig.~$4$, which is -- in my opinion -- the conclusive proof);
\item by studying the unification of relativistic jets from X-ray binaries (XRB) to AGN \cite{FOSCHINI2011,FOSCHINI2012A,FOSCHINI2012B,FOSCHINI2014}.
\end{itemize}

The latter was possible only by adopting the scaling laws by Heinz and Sunyaev \cite{HEINZ2003}, according to which the main driver of the scaling is the mass of the central black hole, while the accretion has much less impact. The scaling is non linear: the jet power scales with $\sim M^{17/12}\sim M^{1.42}$, with slight changes depending on the accretion disk type and the radio spectral index (see Table~1 in \cite{HEINZ2003}). Therefore, jetted NLS1s are the low-mass branch of jetted AGN with radiation-pressure dominated disk, i.e. FSRQs (see \cite{FOSCHINI2011}, Fig. 3). 

Obviously, this generated also a problem of terminology in the classification of jetted AGN, and I promoted new terms based on physical properties like the mass of the central black hole and the cooling of the relativistic electrons of the jet, which in turn depends on the luminosity of the accretion disk \cite{FOSCHINI2017,FOSCHINI2018Z}: FSRQs have high masses and electrons cool very efficiently, so they are called High-Mass Efficient-Cooling AGN (HMEC); jetted NLS1s have the same cooling characteristics, but lower masses, so that they are Low-Mass Efficient-Cooling (LMEC); BL Lac Objects are characterized by inefficient cooling and high masses, so they are named High-Mass Inefficient-Cooling (HMIC). The thresholds dividing the classes are: $M\sim 10^{8}M_{\odot}$ and $L_{\rm disk}\sim 10^{-2}-10^{-3}L_{\rm Edd}$ \cite{FOSCHINI2018Z}. 

Of course, this scenario is not shared by all researchers. There are two main contrapositions: one is to ignore jetted NLS1s (e.g. \cite{PADOVANI2017}), the other is to consider larger masses for the central black hole of NLS1s, equal to FSRQs (see Sect.~\ref{mass}). Among the works of the first type, some are worth mentioning because they include jetted AGN with relatively small masses ($\sim 10^{7}M_{\odot}$), but are not recognized as NLS1s (e.g. \cite{GHISELLINI2009,GHISELLINI2016,GHISELLINI2017,KEENAN2020}). It is not clear if these objects are really FSRQs or misclassified\footnote{There are already known cases where a change of classification was required (e.g. \cite{BERTON2017,JARVELA2020}).} jetted NLS1s or intermediate Seyferts or anything else. What is important is the small mass of the central black hole ($\sim 10^{6-8}M_{\odot}$), independently on the observational appearance. It is necessary to move toward a physics-based classification \cite{FOSCHINI2017}. Today, there could be just a handful of such small-mass AGN, but this is due to the fact that the jet power scales non-linearly with the mass of the central black hole ($P_{\rm jet}\propto M^{17/12}$). Therefore, small-mass AGN emit relatively low-luminosity jets, and, in some cases, their normal luminosity is below the current instruments sensitivities, so that they could be detected only during outbursts (e.g. \cite{LAHTEENMAKI2018}). The advent of {\it Fermi}/LAT, with its superior sensitivity, allowed us to detect the first low-mass jetted AGN\footnote{It is surely not by chance that the first jetted NLS1 to be detected at high-energy $\gamma$ rays was PMN~J$0948+0022$, which lies in the upper end of the NLS1s mass distribution ($M\sim 7.5-15\times 10^{7}M_{\odot}$) \cite{ABDO1,FOSCHINI2015}.}, but it is expected a significant increase of detections as soon as the next generation of instruments will be operative. For example, at radio frequencies, the {\it Square Kilometre Array} (SKA) should detect three orders of magnitudes more jetted NLS1s \cite{BERTONSKA}. 

Works of the second type -- large mass hypothesis -- are based on different methods to estimate the mass: by considering the narrowness of the broad-emission lines of NLS1s as an observational artifact due either to the radiation pressure \cite{MARCONI} or a flat BLR \cite{DECARLI}, by fitting an accretion disk model \cite{CALDERONE}, and by using the black hole-bulge luminosity relationship \cite{LEONTAV2014}. The final result is that the masses increase by about one order of magnitude and jetted NLS1s become common FSRQs (\cite{DAMMANDO2013C,DAMMANDO2019,PALIYA2019,PALIYA2019B}). In addition to the serious flaws already analysed in the previous sections, there is also an unphysical consequence affecting the large-mass hypothesis: the inconsistency between NLS1s, FSRQs, and the electron cooling \cite{FOSCHINI2017}. As known, FSRQs have a photon-rich environment, which means that relativistic electrons of the jet can cool efficiently with long-range multiple collisions with seed photons. BL Lac Objects have a nearby environment poor of photons, which in turn makes it difficult for electrons to cool: the greatest probability to lose energy is via a single head-on collision, transferring most of the electron energy to the seed photon, which in turn jump to TeV energies. Jets from FSRQs have high power, while those from BL Lacs are on the opposite side of the distribution, with low power. This is the well-known {\it blazar sequence} \cite{FOSSATI1998,GHISELLINI1998}. It holds when the masses of FSRQs and BL Lac Objects are distributed on a narrow range ($\sim 10^{8-9}M_{\odot}$), so that the mass scaling does not significantly affect the jet power. Jetted NLS1s have an environment similar to FSRQs, rich of seed photons, but their jet power is comparable with BL Lac Objects \cite{FOSCHINI2015,FOSCHINI2017}. If they have a central black hole with relatively small mass, then the lower jet power is easily explained by the scaling laws of Heinz \& Sunyaev \cite{HEINZ2003}. However, if the mass is in the same range of FSRQs, then it is not possible to explain the measured jet power with the known physics: the electron cooling is based on well-grounded physics and cannot be disputed. Paliya et al. tried to bypass this issue by increasing also the jet power in their SED model ($P_{\rm jet}\sim 10^{45-47}$~erg~s$^{-1}$ \cite{PALIYA2019} to be compared with $P_{\rm jet}\sim 10^{42-45}$~erg~s$^{-1}$ in \cite{FOSCHINI2015}), but then it is no more consistent with the radio luminosity function, where jetted NLS1s are the low-luminosity tail of FSRQs (see Fig.~4 in \cite{BERTON2016}). 

\section{The parent population}
\label{parpop}
In 2012, there were just a handful of jetted NLS1s showing large scale radio structures, but many more were discovered within a few years. Doi et al. were the most prolific hunters of kiloparsec radio structures \cite{DOI2012,DOI2013,DOI2015,DOI2016,DOI2019}, but other significant contributions were either from surveys or individual studies \cite{CACCIANIGA2014,RICHARDS2015,RICHARDS2015B,GU2015,GU2016,CACCIANIGA2017,CONGIU2017,GABANYI2018,BERTON2018,SINGH2018,RAKSHIT2018,GABANYI2019,CHEN2020}. Nevertheless, despite the multiple efforts, the number of jetted NLS1s viewed at large angles, with kiloparsec scale structures, increased to a few tens, is still too low with respect to the expected thousands of objects. This is not so surprising, as already the earliest surveys in 1979 showed that small-mass black holes are generally associated with compact radio structures \cite{MILEY1979}. This can be understood in the framework of the young AGN hypothesis, where the newborn jet had no time yet to build lobes and hot spots, or because its development was hampered by a photon and dust rich interstellar medium (the so-called frustration scenario). A link with CSS/GPS, which are young radio AGN, was proposed since the discovery of PKS~$2004-447$ \cite{OSHLACK,GALLO2006}. 

The key advancement in the search for the parent population was made by Berton et al.: first, they studied the black hole mass distributions among the different candidates, and found that the best option was the steep-spectrum NLS1s population, with some contribution from disk-hosted radio galaxies \cite{BERTON2015}. Then, they studied the luminosity functions and found that the best agreement was with the high-excitation CSS/GPS population (see Fig.~5 in \cite{BERTON2016}). This was the definitive proof of the link between relativistically beamed NLS1s and CSS/GPS. Jetted NLS1s with kiloparsec-scale radio structures are just a few cases, likely the oldest ones before they evolve to FSRQs. This is the cosmological evolution of this type of jetted AGN proposed by Berton et al.: young AGN are NLS1s (relativistically beamed) and high-excitation CSS/GPS (large viewing angle), while FSRQs and high-excitation radio galaxies (HERG) are the old version (see Fig.~3 in \cite{BERTON2017}). The former have small black hole masses, low luminosities, and low jet power, while the latter have large black hole masses, high luminosities, and high jet power. 

There could still be a doubt, that the non-detection of extended radio emission is a matter of instrument sensitivity. This will be settled by forthcoming radio surveys (see Sect.~\ref{persp}).

\section{Outflows}
\label{outflo}
There is another topic where NLS1s break the common paradigm: outflows. It is believed that jets and outflows cannot co-exist, as they are related to different accretion modes: low accretion for jets (radio mode), high accretion for outflows (quasar mode) \cite{GIUSTINI2019}. Again, this proved to be not valid for Seyferts. Earliest detections of outflows in X-ray spectra referred to the jetted Seyferts 3C~111 and 3C~120 \cite{TOMBESI2012O,FUKUMURA2014,TOMBESI2014O}, and were soon followed by detections of outflows in jetted NLS1s -- some of them are $\gamma$-ray detected -- by studying wings and shifts of optical emission lines \cite{KOMOSSA2016,BERTON2016B,KOMOSSA2018B} and X-ray spectra \cite{LONGINOTTI2015,GIROLETTI2017,LONGINOTTI2018,SANFRUTOS2018,COSTANTINI2019,GONZALEZ2020}. These outflows could also interact with the host galaxy on scales much larger than the central parsec, as resulted from millimetre observations of the kinematics of molecular gases \cite{LONGINOTTI2018}. As noted in the Sect.~\ref{hostg}, the host galaxy does not affect the jet formation, but this does not exclude some feedback between the AGN and its host. It would be interesting to set up long and high-cadence multiwavelength campaigns, including optical integral-field spectroscopy, to observe the interplay between jet, outflow, and accretion disk.

\section{Perspectives}
\label{persp}
I think that most of the fundamental questions on jetted NLS1s have been settled: small black hole mass, spiral/disk host galaxy, parent population, young age, relationships with other jetted AGN, particularly with FSRQs\footnote{The terminology issue remains open, but I think it will never close, as we are still using the FRI/FRII divide for radio jets morphology, despite LERG/HERG being much more appropriate.}. Even by recognizing different points of view, other scenarios are severely hampered by strong physical issues. They are implausible, unless new information or new physics will come out in the forthcoming years. Therefore, I think it is possible to say that the study of this type of object is now in a maturity stage. Questions should change accordingly, moving to increase the samples statistics and to search for interesting case studies showing curious anomalies, which could require some intriguing physics to be explained. 

Two major forthcoming facilities could significantly affect the development of the knowledge about these cosmic sources. One is the {\it Square Kilometre Array} (SKA\footnote{\url{https://www.skatelescope.org/}.}), the largest radio telescope with a collecting area of about one square kilometre, as from the name. The construction should start in 2021, but there are already working precursors (e.g. MeerKAT and ASKAP) and pathfinders (e.g. VLA, LOFAR, GMRT, Parkes, e-MERLIN). Extrapolations from the luminosity function indicate that SKA should be able to increase the number of known NLS1s by at least three orders of magnitude \cite{BERTONSKA}. The increase of sample statistics will be essential to consolidate the results obtained to date.

The other facility is the {\it Cherenkov Telescope Array} (CTA\footnote{\url{https://www.cta-observatory.org/}.}), which will work at GeV-TeV energies. Also in this case, there are working prototypes, but the array construction should end by 2025. Romano et al. performed simulations with increasing details to understand the potentiality of CTA, and found some jetted NLS1s worth observing, particularly during outbursts \cite{ROMANO2018A,ROMANO2018B,ROMANO2018C}. In some cases, CTA should be able to distinguish the location where $\gamma$ rays are generated, if within the BLR or outside it, around the molecular torus \cite{ROMANO2020}. Vercellone et al. studied the possibility to distinguish the shape of the $\gamma$-ray spectrum, if a log-parabola or a power law model \cite{VERCELLONE2020}. This can have impact in understanding the mechanisms at work to accelerate particles to the energies of cosmic rays. 

Since the low-energy threshold of CTA is $\sim 20$~GeV, it is important that this ground-based facility will be combined with a space telescope operating at MeV-GeV energies: if {\it Fermi} will be no more available, it is necessary to design its successor. There are some projects under development (e.g. {\it e-ASTROGAM} \cite{DEANGELIS}, {\it AMEGO}\footnote{\url{https://asd.gsfc.nasa.gov/amego/}.}), but presently none is for sure.

In the X-rays energy band ($0.3-10$~keV), great discoveries are expected from the all-sky survey of {\it eROSITA}\footnote{\url{https://www.mpe.mpg.de/eROSITA}}, launched on 2019 July 13, that just started releasing wonderful images while I was writing the present review. Another satellite worth noting is NASA {\it Imaging X-ray Polarimetry Explorer} (IXPE\footnote{\url{https://ixpe.msfc.nasa.gov/}.}), to be launched in April 2021 and dedicated to the X-ray polarimetry in the $2-8$~keV band. Some notes on the expected polarization properties of NLS1s \cite{FOSCHINI2016} were prepared for {\it XIPE}, an ESA competitor that was not accepted, but can be useful also for the {\it IXPE} case, by taking into account that the latter is a smaller version of the former (effective area at $3$~keV: $700$ vs $1100$~cm$^2$). A satellite for a quick follow-up at ultraviolet/X-rays frequencies to replace {\it Swift} is also currently missing.
 
Optical/infrared spectroscopy is one of the most important observational window in this research field. Although there are many ways to identify an AGN, the optical spectroscopy is still the best one, the richest of information. The {\it Sloan Digital Sky Survey} (SDSS\footnote{\url{https://www.sdss.org/}.}) played a key role in the identification of NLS1s and the generation of flux-limited samples, with its impressive wealth of high-quality spectra freely available. The forthcoming SDSS-V\footnote{\url{https://www.sdss.org/future/}.} is a guarantee of continuity. However, a similar project is missing in the Southern Hemisphere.  

\section{All that jets}
In the previous sections, I have reviewed what I think are the main works in this research field published between $2012$ and $2020$. Obviously, this is a personal selection and is biased according to my preferences, interests, and pet theories. Before concluding, I would like to list at least all the papers (articles and proceedings; excluding presentations with only slides or abstracts, and electronic circulars) I have found in the available literature. This should balance my biases and gives an almost complete list of the work done in this field. The following list is coarsely divided depending on the topic.

\subsection{General reviews}
A conference on NLS1s was held in Padova in 2018, the third specific on NLS1s, after the first one held in 1999 ({\it Joint MPE-AIP-ESO workshop on Observational and Theoretical Progress in the Study of Narrow-Line Seyfert 1 Galaxies}, December 8-11, 1999, Bad Honnef, Germany, \cite{WORKSHOP1999}), and the second one held in 2011 ({\it Narrow-Line Seyfert 1 Galaxies and their place in the Universe}, April 4-6, 2011, Milano, Italy, \cite{WORKSHOP2011}). In the proceedings of the 2018 Padova conference \cite{WORKSHOP2018}, there are many invited reviews worth mentioning: Komossa on the multiwavelength properties \cite{KOMOSSA2018}, Lister on the radio characteristics \cite{LISTER2018}, Gallo about the X-ray emission \cite{GALLO2018}, and Czerny to put NLS1s in the wide context of the quasar main sequence \cite{CZERNY2018}.

It is also worth reminding the conference {\it Nuclei of Seyfert galaxies and QSOs - Central engine \& conditions of star formation} (November 6-8, 2012, Bonn, Germany), which included sessions dedicated to NLS1s and jets \cite{WORKSHOP2012}. This is also the workshop for which I wrote the 2012 review, which is the starting point of the present essay \cite{FOSCHINI2}.

With reference to other reviews, there are many contributions by D'Ammando et al. (including also some papers where NLS1s were only a section of a broader review on relativistic jets, \cite{DAMMANDO2013B,DAMMANDO2013C,DAMMANDO2013D,DAMMANDO2014,DAMMANDO2016,DAMMANDO2017B,DAMMANDO2019}), Ermash and Komberg \cite{ERMASH2013} (general review on physical and observational properties), Boller (on history of X-ray detections \cite{BOLLER2014}), and Paliya (on gamma-ray NLS1s \cite{PALIYA2019}). More general reviews including sections dedicated to NLS1s were written by Rieger (gamma rays from non-blazar AGN, \cite{RIEGER2017}) and Hada (high-angular resolution radio observations, \cite{HADA2019}).

\subsection{Case studies, Individual objects, Anomalies}
Most of the works were focused on $\gamma$-ray detected NLS1s, either with the search for new detections or with multiwavelength campaigns on individual objects. Recent lists of $\gamma$-NLS1s can be found in \cite{ROMANO2018C,PALIYA2019B}.

1H~$0323+342$ is the most studied object, being also the closest NLS1 detected at MeV-GeV energies. At $z=0.063$, one milliarcsecond is equivalent to $\sim 1.2$~pc, so that it was possible to study its radio morphology with great details \cite{WAJIMA2014,WAJIMA2014B,FUHRMANN2016,DOI2018B,HADA2018,KOVALEV2020,DOI2020}. Particularly, a transition of the jet shape (from parabolic to conical, as the distance from the central black hole increases) was observed, similar to that in M87 \cite{DOI2018B,HADA2018,KOVALEV2020,DOI2020}. This confirmed also the scaling laws by Heinz and Sunyaev \cite{HEINZ2003}, because it was possible to overlap the jet morphologies of the two objects with the proper scaling \cite{FOSCHINI2019}. Other studies at different wavelengths are \cite{TAKAHASHI2012,TIBOLLA2013,ITOH2014,YAO2015,MILLER2017,WANG2017,ARRIETALOBO2017,BOISSON2018,KYNOCH2018,GHOSH2018,MUNDO2020}.

PMN~J$0948+0022$ ($z=0.585$), the first NLS1 to be detected at high-energy $\gamma$ rays, continued to be observed at any frequency \cite{FOSCHINI2012,MAUNE2013,EGGEN2013,DAMMANDO2014Z,BHATTA2014,SUN2014,DAMMANDO2015Z,ZHANG2017}. It is worth noting the observation of strong variability at optical frequencies ($R_{\rm C}$ filter) with time scale of the order of minutes ($2.3-3.0$~min) and strong polarization ($36\pm 3$\%) \cite{ITOH2013}, which is again a further proof of the small mass of the central black hole. 

Other $\gamma$-detected NLS1s with dedicated works are:

\begin{itemize}
\item SBS~$0846+513$ ($z=0.584$) \cite{DAMMANDO2012,DAMMANDO2013Z,PALIYA2016Z};
\item PKS~$1502+036$ ($z=0.408$) \cite{PALIYA2013,DAMMANDO2013Z2,BULGARI,PALIYA2016Z2,DAMMANDO2016Z2};
\item PKS~$2004-447$ ($z=0.24$) \cite{PALIYA2013,BALDI,ORIENTI2015,KREIKENBOHM2016,SCHULZ2016};
\item FBQS~J$1644+2619$ ($z=0.145$) \cite{DAMMANDO2015ZZ,OIGLE2017,DAMMANDO2017,LARSSON2018};
\item 3C~286 ($z=0.850$) \cite{BERTON2017};
\item B3~$1441+476$ ($z=0.703$) and GB6~B$1303+5132$ ($z=0.785$) \cite{LIAO2017};
\item TXS~$2116-077$ ($z=0.26$) \cite{YANG2018,PALIYA2020,JARVELA2020};
\item PKS~J$1222+0413$ ($z=0.966$) \cite{YAO2015B,KYNOCH2019};
\item TXS~$0943+105$ ($z=1.004$) \cite{YAO2019}.
\end{itemize}

The two latter NLS1s are particularly interesting: they are around $z\sim 1$, a distance where it starts to be possible to check the validity of the hypothesis of the recent cosmological birth of NLS1s. Previous surveys based only on optical spectroscopy made it possible detections at $z\lesssim 0.8$ \cite{ZHOU2006}, but this due to the fact that at $z\gtrsim 0.8$, the H$\beta$ line ($\lambda_{\rm rest}=4861$\AA) shifts from optical to infrared wavelengths. Therefore, the lack of NLS1s at $z\gtrsim 0.8$ could simply be an observational bias. Yao et al.'s work on the SDSS Baryon Oscillation Spectroscopic Survey (BOSS \cite{DAWSON2013}, whose wavelength range extends to $10140$\AA\, making it possible to detect H$\beta$ up to $z\sim 1.08$), confirmed that there are NLS1s at $z\gtrsim 0.8$ and the previous limit was an observational artifact. Additional work is necessary to find more NLS1s at $z\sim 1$ and even beyond, in order to test the validity of the young AGN hypothesis.

In addition to the confirmed $\gamma$-ray NLS1s, there are also other objects candidate to be $\gamma$-ray emitters, with one not yet confirmed detection: SDSS~J$164100.10+345452.7$ ($z=0.164$) \cite{LAHTEENMAKI2018}, RX~J2314.9+2243 ($z=0.169$) \cite{KOMOSSA2015,MILLER2017B}, and IRAS~$20181-2244$ ($z=0.185$) \cite{MILLER2017B,BERTON2019B}. It is also worth noting that the search for NLS1s extended to anomalous objects, with uncertain classification. It was suggested that the $\gamma$-detected quasar B2~$0954+25$A ($z=0.712$) \cite{CALDERONE2012,ARRIETALOBO2017} could be a transition object between FSRQs and NLS1s. The H$\beta$ shape is significantly distorted and the proposed classification as NLS1 is based in considering only the symmetric component. 

Other jetted NLS1s, although not detected at high-energy gamma rays, have been studied in details: 

\begin{itemize}
\item SDSS~J$110006.07+442144.3$ ($z=0.840$) \cite{TANAKA2014,GABANYI2018};
\item RX~J$1633.3+4719$ ($z=0.116$) \cite{MALLICK2016};
\item Mrk 1239 ($z=0.020$) \cite{DOI2015};
\item SDSS J$103024.95+551622.7$ ($z=0.435$) \cite{RAKSHIT2018,GABANYI2019};
\item IRAS~$17020+4544$ ($z=0.060$) \cite{LONGINOTTI2015,GIROLETTI2017,LONGINOTTI2018,SANFRUTOS2018,GONZALEZ2020} (this is particularly interesting, as it shows the coexistence of a relativistic jet and ultrafast outflows);
\item Mrk 783 ($z=0.067$) \cite{CONGIU2017};
\item SDSS~J$143244.91+301435.3$ ($z=0.355$) \cite{CACCIANIGA2014,CACCIANIGA2017};
\item PKS~$0558-504$ ($0.137$) \cite{GLIOZZI2013,GHOSH2016}.
\end{itemize}

It is worth concluding this subsection with an overview on jetted Seyferts and disk-hosted radiogalaxies. For the sake of simplicity, I refer only to those detected at high-energy gamma rays:

\begin{itemize}
\item IC~310 ($0.019$) \cite{KADLER2012,ALEKSIC2014,ALEKSIC2014B,AHNEN2017,GHOSAL2018,GRAHAM2019};
\item III Zw 2 ($0.089$) \cite{CHEN2012,LIAO2016,GONZALEZ2018,CHAMANI2020};
\item 3C~120 ($z=0.033$) \cite{AGUDO2012,COWPERT2012,POZO2012,LOHFINK2013,LIU2014,KOLLATSCHNY2014,POZO2014,SAHAKYAN2015,TANAKA2015,LIU2015,CASADIO2015,RAMOLLA2015,JANIAK2016,DONS2016,CASADIO2016,TOMBESI2017,ZARGARYAN2017,RANI2018,MARSCHER2018,RAMOLLA2018,WILKINS2019,HLABATHE};
\item 3C~111 ($z=0.048$) \cite{GROSSBERGER,BELL,GRANDI2012,TOMBESI2012,DEJONG2012,TOMBESI2013,CLAUTICE2016,TOMBESI2017B,BEUCHERT2018}.
\end{itemize}

I would like also to remind that the literature collected here refers to the time interval $2012-2020$.

\subsection{Sample study, surveys}
In addition to studies on individual objects, many studies on samples (from just a few to many sources) have been published. The largest multiwavelength sample (292 NLS1s with radio counterpart) was studied by J\"arvel\"a et al. \cite{JARVELA2015}. Another sample made of 42 jetted NLS1s was studied at all the available frequencies \cite{FOSCHINI2015}, while the broad-band spectra of a sample of 16 jetted NLS1s detected at $\gamma$ rays was fit by using a one-zone leptonic model \cite{PALIYA2019}. Other studies based on simple radiative models applied to broad-band spectra of NLS1s and comparison with other types of blazars can be found in these works \cite{ZHANG2013,SUN2015,ZHANG2015,YANG2015,ZHU2016}. The SED is not the only way to study the jet physics: simpler, but not simplistic, methods are based on some direct relationships between observed and physical quantities, like, for example, the radio core luminosity with the jet power \cite{BLANDFORD}. Examples of this type of study are \cite{FOSCHINI2011,FOSCHINI2012A,FOSCHINI2012B,GAROFALO2013,FOSCHINI2014,YAO2018,FAN2019,CHENY2019,LIU2020}. 

Focusing on specific frequencies, the search for new detections at high-energies $\gamma$ rays continued as the {\it Fermi} dataset increases \cite{MILLER2017B,PALIYA2018}, as well as monitoring or reanalysis of LAT data \cite{PALIYA2015Z,BAGH2018,SAHAKYAN2018}. As already noted, presently the sample of firm $\gamma$-ray detections is composed of about two dozens of NLS1s \cite{ROMANO2018C,PALIYA2019B}, although reclassification of known $\gamma$-detected AGN has a great potentiality (e.g. \cite{BERTON2017}).

Several studies on samples at radio frequencies have been done: \cite{ORIENTI2012,ANGELAKIS2012,DOI2012,DOI2013,PANESSA2013,ANGELAKIS2013,KHARB2014,RICHARDS2015,RICHARDS2015B,ANGELAKIS2015,GU2015,GU2016,DOI2016,KARA2017,GU2017,LAHTEENMAKI2017,BERTON2018,LAHTEENMAKI2018,SINGH2018,FAN2020}. Works worth noting are those based on 15~GHz VLBA observations of the MOJAVE Program\footnote{Monitoring Of Jets in Active galactic nuclei with VLBA Experiments, \url{https://www.physics.purdue.edu/MOJAVE/}.}, which made it possible to study the kinematic of the jet components, and hence to measure superluminal $\beta_{\rm app}$ (up to $\sim 11c$), and the polarization \cite{LISTER2016,HODGE2018,LISTER2019}. It is also worth mentioning the work by Chen et al. \cite{CHEN2020} on the first sample of NLS1s in the Southern Hemisphere.

Other studies at infrared/optical wavelengths \cite{PALIYA2013Z,CACCIANIGA,KAWAGUCHI2015,LIU2016,LIU2017,KSHAMA2017,RAKSHIT2017,RAKSHIT2018B,ANGELAKIS2018,OJHA2018,OJHA2019,OJHA2020} and at X-rays \cite{RANI2017,BAGH2018,BERTON2019C} have been done.

\subsection{Other topics}
There are other specific topics deserving some attention. J\"arvel\"a et al. \cite{JARVELA2017Z} studied the large scale environment of NLS1s, and found that NLS1s are placed in less dense regions of the Universe, at odds with broad-line Seyferts, thus confirming their young age. However, as the radio loudness increases, the density of the large scale environment increases. Perhaps, this is an indication that NLS1s could be a mixed bag containing different types of objects.

Sbarrato et al. \cite{SBARRATO2018} studied the optical and radio properties to understand the geometrical orientation of different types of NLS1s (radio silent, quiet, and loud). They found that the fraction of jetted NLS1s is larger than what found in other classes of AGN.

There are other attempts to study the scaling laws of jetted AGN, including NLS1s \cite{LIODAKIS2017,GARDNER2018}, but they were not able to follow the Heinz and Sunyaev \cite{HEINZ2003} relationships. 

Some numerical simulations have also been done, with specific reference to the jet launching from slim disks \cite{CAO2014,MA2014,MA2014B}.

\section{Final remarks}
While taking into account conflicting opinions, the emerging scenario is the following: jetted NLS1s are AGN with relatively small masses of the central black hole ($\sim 10^{6-8}M_{\odot}$), disks with high accretion luminosity ($\gtrsim 0.01-0.001L_{\rm Edd}$), jets with relatively average-low power ($\sim 10^{42-45}$~erg~s$^{-1}$), host galaxies of different types (disk, spiral, ellipticals, with bars or pseudobulges, with ongoing or past interactions). From the observational point of view, jetted NLS1s are different from blazars and radiogalaxies, but from the physical point of view, the central engine and the jet are the same, and follow the scaling laws of Heinz and Sunyaev \cite{HEINZ2003}, the same laws that rule the scaling toward jetted XRBs. In the framework of the unification of relativistic jets, NLS1s are essential, because they are the low-mass branch of AGN (particularly FSRQs), in parallel to the accreting neutron stars for XRBs. 

Two fundamental points have been settled by jetted NLS1s with respect to the generation of relativistic jets: (i) there is no mass threshold (ii) the host galaxy does not matter. This should not be surprising, because we knew since long time the existence of relativistic jets from XRBs, where the compact object has mass much smaller than an AGN and the environment is quite different. It is worth noting that although the host galaxy does not affect the jet generation, this does not exclude some feedback with the central engine via large scale outflows. 

I think that presently this research field has reached a maturity stage, but we should be aware that we have just grazed the tip of an iceberg. When SKA will start observations, the number of jetted NLS1s should extremely increase. Large surveys will be essential to consolidate the current results, but must be associated with the monitoring of peculiar objects, which in turn are conducive to new and unexpected discoveries (e.g. \cite{LAHTEENMAKI2018}). These two lines of research should result in important and new information on the generation, development, and stop of jets, the duty cycle, the connection with the accretion disk and the outflows, the feedback with the host galaxy. Time-resolved multiwavelength astrophysics will be the fundamental asset. It is now evident that classifications based on single observations made within a restricted range of frequencies are obsolete. Extragalactic objects change in time, even on human time scales. 

Among the forthcoming observing facilities, something is still missing at high energies. CTA will surely be able to give important contributions to the jet physics, particularly on the mechanisms of particle acceleration, but its low-energy threshold of $\sim 20$~GeV is too high for a stand-alone facility. A satellite operating at MeV-GeV to replace {\it Fermi} is needed. There are interesting projects, but currently with uncertain result. The ultraviolet/X-ray band is also missing a replacement for the {\it Swift} satellite, to operate a quick follow-up and -- most important -- to be able to satisfy almost all the requests. 

I would like to conclude with a few words on terminology. It is very difficult to change words habits, and we scientists are not different from other human beings. Nevertheless, we must do a major effort to change. It seems there is a serious problem to move from merely observational classifications to others with physical content. This is not new: already in 2012, Gaskell complained about this bad habit \cite{GASKELL}. It is true that words do not affect the physics and chemistry of physical objects, but they structure our way to think (e.g. \cite{FOSCHINI1Z,FOSCHINI2Z}). If we will not be able to update our words, our terminology, and our classifications according to new discoveries, we cannot advance our level of knowledge. 

\vspace{6pt} 

\section*{Acknowledgments}
I would like to warmly thank (in alphabetical order) drs. Marco Berton, Stefano Ciroi, Pat Romano, and Stefano Vercellone for their critical reading of the draft of this essay, precious advices, and kind help. Thanks also to the three anonymous referees for their useful comments.

\end{document}